\documentclass[reprint,showpacs,preprintnumbers,amsmath,amssymb]{revtex4-1}

\usepackage[T1]{fontenc}
\usepackage{color}
\usepackage[latin1]{inputenc}
\usepackage{graphicx}
\usepackage{bm}
\usepackage{epsfig}
\usepackage{rotating}

\usepackage{dcolumn}
\usepackage{bm}

\begin{document}


\title{Capillary and winding transitions in a confined cholesteric liquid crystal}
       
\author{Daniel de las Heras}
\email{delasheras.daniel@gmail.com}
\affiliation{Theoretische Physik II, Universit\"at Bayreuth, D-95440, Bayreuth, Germany}

\author{Enrique Velasco}
\email{enrique.velasco@uam.es}
\affiliation{Departamento de F\'{\i}sica Te\'orica de la Materia Condensada,
Instituto de F\'{\i}sica de la Materia Condensada (IFIMAC) and Instituto de Ciencia de Materiales Nicol\'as Cabrera,
Universidad Aut\'onoma de Madrid,
E-28049, Madrid, Spain}

\author{Yuri Mart\'{\i}nez-Rat\'on}
\email{yuri@math.uc3m.es}
\affiliation{Grupo Interdisciplinar de Sistemas Complejos (GISC), Departamento de Matem\'aticas, Escuela
Polit\'ecnica Superior, Universidad Carlos III de Madrid, Avenida de la Universidad 30, E-28911, Legan\'es, Madrid, Spain}

\date{\today}

\begin{abstract}
We consider a Lebwohl-Lasher model of chiral particles confined in a planar cell (slit pore) with different boundary conditions, and solve it
using mean-field theory. The phase behaviour of the system with respect to temperature and pore width is studied.
Two phenomena are observed: (i) an isotropic-cholesteric transition which exhibits an oscillatory structure with respect to pore width, and (ii)
an infinite set of winding transitions caused by commensuration effects between cholesteric pitch and pore width. The latter transitions 
have been predicted and analysed by other authors for cholesterics confined in a fixed pore and subject to
an external field promoting the uniaxial nematic phase; here we induce winding transitions solely from geometry by changing the pore width 
at zero external field (a setup recently explored in Atomic-Force Microscopy experiments). In contrast with previous studies, we obtain 
the phase diagram in the temperature vs pore width plane, including the isotropic-cholesteric transition, the winding transitions and
their complex relationship. In particular, the structure of winding transitions terminates at the capillary isotropic-cholesteric transition 
via triple points where the confined isotropic phase coexists with two cholesterics with different helix indices.  
For symmetric and asymmetric monostable plate anchorings the phase diagram are qualitatively
similar. 
\end{abstract}

\keywords{cholesteric, confined, liquid crystals, phase transitions}

\maketitle

\section{Introduction}

Cholesteric phases of liquid crystal materials are important from fundamental and technological points of view \cite{deGennes,Blinov}, 
and a lot of effort is still
being expended in their study. Interest in cholesterics stems from the existence of a direction, perpendicular to the nematic director, about which the latter
winds up with a specific periodic length, the pitch, causing the material to be optically active at optical wavelengths. The pitch sensitivity to temperature and
external fields, and the ensuing varying reflectivity properties, can be exploited in applications. Also, temperature variations of the pitch can be
suppressed by different techniques, e.g. by mixing two chiral dopants with opposite temperature dependence in their pitch \cite{koreanos}.

Cholesteric materials continue to be of interest, in particular, as the basis for the design of the twisted nematic cells of liquid-crystal-based
devices. Many experimental and theoretical works have been devoted to this problem. Still many basic issues remain unexplored in the field of
capillary phenomena in cholesteric materials. For example, the first theoretical study of the isotropic-cholesteric interface is very recent \cite{NelsonNunoSoftMatter}.

Most theoretical studies of cholesterics have been based on the Landau-de Gennes model, especially in connection with the existence of the blue phases (see e.g.
\cite{Grebel}). Another fruitful approach is the chiral Lebwohl-Lasher model \cite{LL1,LL2}, which has been studied from various perspectives. 
The bulk properties of the chiral Lebwohl-Lasher model were explored by van der Meer et al. \cite{vanderMeer} using mean-field theory. 
The most striking conclusion was that the pitch was independent of temperature, and that higher-order terms in the interactions were necessary to
restore a temperature-dependent pitch. The model was first simulated by Saha et al. \cite{Saha0} using the Monte Carlo technique and a planar
rotor model for the particles. Discrepancies were found initially concerning the independency of the pitch with temperature. Later 
simulations by Saha and Saha \cite{Saha} of the full model (without orientational restrictions) and of the planar rotor model by 
Luckhurst et al. \cite{Luckhurst} confirmed the mean-field predictions. The latter authors obtained a continuous isotropic-cholesteric 
transition, and explained the temperature-independent feature as a result of the existence of a helical transformation, 
suggested by T. J. Sluckin, that removes the chiral term in the energy. Note that, in more realistic models with density fluctuations, one expects the pitch
to depend on the temperature due to variations in density and order parameter with temperature.

Other authors have used the model to explore different features of chiral phases. 
Memmer and Janssen \cite{first} studied the influence of the fourth-order terms on the pitch, using simulation and self-determined 
boundary conditions. In agreement with van der Meer et al. \cite{vanderMeer}, temperature-dependency of the pitch was restored.
A helix inversion at a specific temperature was obtained by these authors with the same fourth-order interaction potential \cite{Memmer1}, 
again in accord with the mean-field calculation of van der Meer et al. \cite{vanderMeer}.
Memmer and Janssen \cite{Memmer0} also studied the induction of a helical twist by mixing chiral dopant and a guest achiral system, 
a common procedure in experiments and applications to optimise device performance.

The chiral Lebwohl-Lasher model has also been investigated in confined geometry. This is a relevant model to study twisted liquid-crystal 
displays. These systems operate with the help of an external field. To cite a representative example of a simulation analysis,
Memmer and Fliegans \cite{Fliegans} studied a chiral Lebwohl-Lasher model confined by two parallel plates and subject to a constant external
field normal to the plates. A corresponding Fre\'edericksz transition between twisted and uniform director arrangements was characterised.
An external electric field can be used to control the pitch in a bulk cholesteric sample. In the case of a cholesteric confined between
two plates a fixed distance apart (pore width), there is a commensuration issue, and the variation of the number of pitch lengths
can only occur in discrete jumps involving structures with different number of helix indices (number of half turns of the director about the 
helix direction) \cite{Dreher}. These variations can be induced by temperature or by an electric field. Previous and more recent works have 
analysed the problem using simple elastic models \cite{Kedney1,Kedney2,Belyakov1,Belyakov2,Oswald,Seidin,McKay1,McKay2}.
There are various experimental works that deal with these problems \cite{Regaya,Patel,Lee,Sprang,Niggemann,Schlangen,Yip,Smalyukh,Yoon}.

Our purpose in the present work is to analyse commensuration effects in confined chiral liquid crystals when the agent that changes
the number of pitch lengths is not temperature or an external field, but the pore width. Therefore, we are inducing the 
winding transitions by geometry alone. An important feature of our theoretical technique is that 
order parameters and a free-energy functional are used; therefore the nematic and cholesteric order are analysed consistently
as a function of temperature, and the confined order-disorder transition and structural phenomena in the slit pore can be followed and
described collectively. Also, we consider a free-orientation model, so that the main axes of particles are not restricted to lie on planes
parallel to the confining plates.

To help focus our work, we can think of a material consisting of a mixture of
two chiral substances such that the pitch temperature dependence is suppressed \cite{koreanos}. This material can be placed
in a cell with strong anchoring conditions, such that the pore width can be controlled, as in an Atomic-Force Microscopy (AFM) 
experiment; indeed such a setup has been used to identify capillary nematization transitions \cite{eslovenos} and, 
more recently, even to obtain the pitch in cholesteric materials with high accuracy \cite{eslovenos1}. The control parameter
is $h/p$, where $h$ is the pore width and $p$ the cholesteric pitch; this parameter dictates the number of cholesteric periods or helix index
in the cell, and its variation will induce transitions between cholesteric
slabs of different numbers of half-turns of the director (winding transitions). As in the AFM experiments \cite{eslovenos1}, an infinite collection of 
winding transitions are obtained. Using the chiral Lebwohl-Lasher
model, we analyse these transitions in detail, locate the capillary isotropic-cholesteric phase transition as a function of the pore width, and
establish the connection between this transition and the winding transitions, a relationship that previous studies have not addressed. 
An interesting topology in the temperature vs. pore width
phase diagram is observed. The cases of monostable symmetric and asymmetric surface alignments
are considered, and the differences between the two cases discussed.

\section{Theory}

\begin{figure}
\begin{center}
\includegraphics[width=0.95\linewidth,angle=0]{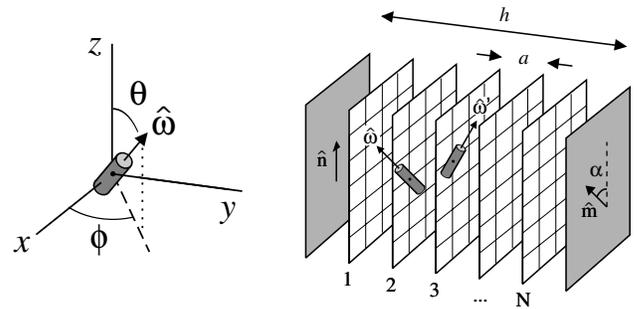}
\caption{Setup for the Lebwohl-Lasher cell: a simple-cubic lattice sandwiched between two parallel plates normal to the [100] direction.
Planes parallel to the plates are indicated by an index from 1 to $N$. The pore width is $h=(N+1)a$, where $a$ is the lattice parameter.
The favoured alignment at each plate is given by unit vectors $\hat{\bm n}$
and $\hat{\bm m}$, respectively, with $\alpha$ the angle between the two vectors. All lattice sites are occupied by one particle with orientation 
$\hat{\bm\omega}$. Cartesian axes used (lab reference frame) are shown at left, along with the polar angles associated with $\hat{\bm\omega}$
(note that the $z$ axis lies in the plane of the plates whereas the $y$ axis is normal to the plates).}
\label{fig1}
\end{center}
\end{figure}

The model used to explore the confined cholesteric material is the chiral Lebwohl-Lasher model. This model has some advantages over
the popular Landau-de Gennes or elastic models used in previous studies. The most important of them is the fact that, in contrast to 
the latter approaches, the model has a microscopic nature
and the interaction potential appears explicitly. Different terms in the interaction potential can be tuned to model specific features of
real molecules. By contrast, the Landau-de Gennes or elastic theories are meso- and macroscopic in nature, aim at a description at larger length 
scales, and become the logical choices when the nematic director is spatially deformed on a scale much larger than the molecular 
scale,
or surfaces induce the presence of defects.  
These scales are difficult to access using a microscopic approach due to computational limitations.

In the Lebwohl-Lasher model, each site of a simple-cubic lattice of lattice parameter $a$ is filled with a uniaxial particle described by an 
orientation $\hat{\bm\omega}=(\theta,\varphi)$. Particles interact via a nearest-neighbour potential that depends on the two particle
orientations $\hat{\bm\omega}$, $\hat{\bm\omega}'$ and possibly also on the relative position vector $\hat{\bm r}$.
In the chiral Lebwohl-Lasher model, the potential energy is taken as
\begin{eqnarray}
&&\Phi(\hat{\bm\omega},\hat{\bm\omega}',\hat{\bm r})\nonumber\\
&&=-\epsilon P_2\left(\hat{\bm\omega}\cdot\hat{\bm\omega}'\right)-
\kappa\left[\left(\hat{\bm\omega}\times\hat{\bm\omega}'\right)\cdot\hat{\bm r}\right]
\left(\hat{\bm\omega}\cdot\hat{\bm\omega}'\right),
\label{interac}
\end{eqnarray}
where $P_2(x)$ is the second-order Legendre polynomial.
The case $\kappa=0$ corresponds to the standard, achiral Lebwohl-Lasher model.

These terms are the first scalar and pseudo-scalar terms of a general expansion in rotational invariants of the interaction potential associated 
with two uniaxial particles \cite{vanderMeer,Osipov}. The first, Maier-Saupe-like, term favours arrangements where particles are parallel. The second 
term takes account of the chirality of the 
interaction (in the following the chiral strength will be measured by $\kappa^*=\kappa/\epsilon$).

Particles in the Lebwohl-Lasher lattice model can be interpreted as single molecules, but Memmer and Fliegans \cite{Fliegans}
have also interpreted each particle on the lattice as a director representing
a cluster of neighbouring molecules. In our case the different interpretations are not relevant considering the qualitative level of 
our work. 

The setup for the confined Lebwohl-Lasher model is shown in Fig. \ref{fig1}: the lattice is sandwiched between
two parallel plates normal to the [010] direction. The $z$ axis is chosen parallel to the plates, along one of the two
equivalent crystalline axes.
In the spirit of mean-field theory, all particles pertaining to the same plane
parallel to the surfaces (labelled with the index $i=1,2,\cdots,N$; see Fig. \ref{fig1}) will be taken as equivalent. 
Particles in planes $1$ and $N$ interact with their closest plate via the surface potentials $\Phi_1(\hat{\bm\omega})$ and
$\Phi_2(\hat{\bm\omega})$. The following cases are considered:
\begin{itemize}
 \item{\bf Symmetric plates}. Both plates have one easy axis, contained in the plane of the plates, and the two easy axes are identical, $\hat{\bm n}=\hat{\bm m}=(0,0,1)$ (see left panel
 of Fig. \ref{fig1} for our choice of reference axes). The corresponding surface potential energies are monostable, and give contributions  
 $\Phi_i(\hat{\bm\omega})=\epsilon_sP_2(\hat{\bm n}\cdot\hat{\bm\omega})$, $i=1$ and $2$, only from the particles adjacent to the two plates.
 \item{\bf Asymmetric plates}. The easy axes of the two plates form an angle of $\alpha=90^{\circ}$, i.e. $\hat{\bm n}=(0,0,1)$ and
 $\hat{\bm m}=(1,0,0)$. Contributions from particles adjacent to the plates are 
 $\Phi_1(\hat{\bm\omega})=\epsilon_sP_2(\hat{\bm n}\cdot\hat{\bm\omega})$ and
 $\Phi_2(\hat{\bm\omega})=\epsilon_sP_2(\hat{\bm m}\cdot\hat{\bm\omega})$, respectively. Note that both potential energies are monostable,
 and that their strengths are identical; only the easy axis is different.
\end{itemize}
All surface potentials used
are built from second-order Legendre polynomials. This is an advantage since 
then all interactions turn out
to be quadratic in the unit vector $\hat{\bm\omega}$, which simplifies the theoretical treatment. 
We take $\epsilon_s=\epsilon$ (we later argue that this choice leads to strong anchoring conditions).
In the following we use the Maier-Saupe parameter $\epsilon$ as an energy scale. We therefore have the two parameters 
$kT/\epsilon$ and $\kappa/\epsilon$ as energy parameters, the length parameter being $h/p$ (both $h$ and $p$ are measured in lattice unit lengths $a$,
 and this is taken as the length scale in the model).

 \begin{figure*}
\includegraphics[width=0.80\linewidth,angle=0]{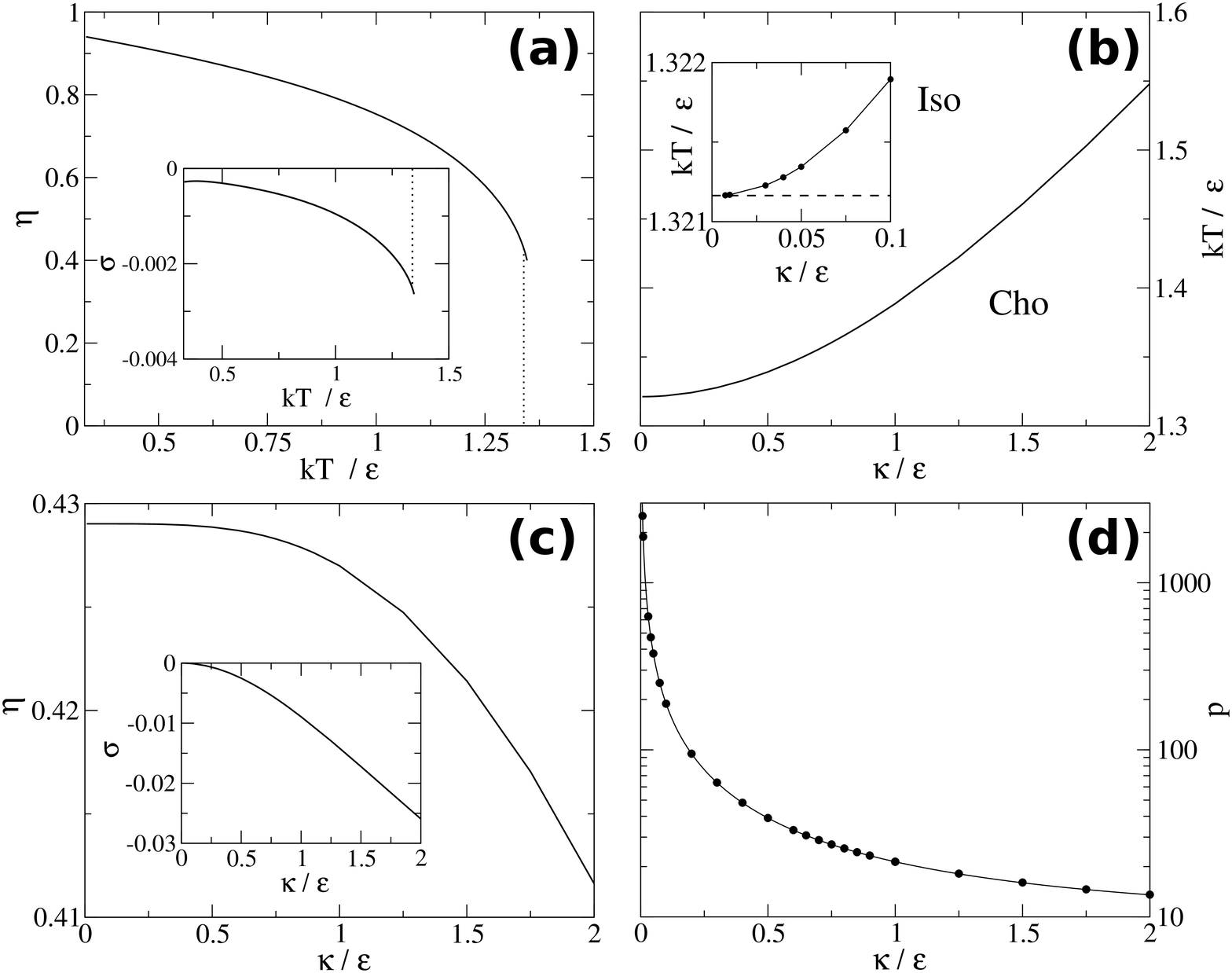}
\caption{(a) Uniaxial order parameter $\eta$ and biaxial order parameter $\sigma$ (inset) 
as a function of the reduced temperature $kT / \epsilon$ for $\kappa/\epsilon=0.5$. The 
vertical dotted line indicates the isotropic-cholesteric first order phase transition. 
The curves are qualitatively the same for other values of $\kappa$. (b) Bulk phase diagram in the reduced temperature-$\kappa$ plane. 
The black-solid line is the isotropic-cholesteric binodal. The nematic state is only stable in the limit $\kappa\rightarrow0$. 
The inset is a zoom of the small $\kappa$ region, showing the (unstable) isotropic-nematic transition (red line), which does not depend on $\kappa$. (c) Uniaxial order 
parameter $\eta$ and biaxial order parameter $\sigma$ (inset) as a function of $\kappa$ at the isotropic-cholesteric phase transition. (d) Pitch $p$ as a function of the chirality parameter $\kappa$ at 
the isotropic-cholesteric phase transition. The symbols are results obtained via minimization of the free energy and the line is the approximation 
$p=4\pi(\arctan\frac{2\kappa}{3\epsilon})^{-1}$.}
\label{fig2}
\end{figure*}

Due to the inhomogeneity of the problem, the orientational order will vary from plane to plane. Since in-plane fluctuations are neglected,
the orientational distribution function will only depend on the plane index $i$, i.e.
$f_i(\hat{\bm\omega})$, with $i=1,\cdots, N$. 
The statistical mechanics of the model is solved using a variational principle on the mean-field free-energy functional (for details
see Marguta et al. \cite{Marguta}).
The functional, $F[f_i]$, has contributions from entropy, from interactions between particles, and
from interactions between particles and the plates. The energy has contributions from the four particles in the same layer,  
from the two particles on the two neighbouring layers, and also from
the external potential. The complete free energy functional is:
\begin{widetext}
\begin{eqnarray}
\frac{F[\{f_i\}]}{\epsilon M}&=&\frac{kT}{\epsilon}
\sum_{i=1}^N\int d\hat{\bm\omega}f_{i}(\hat{\bm\omega})
\log{\left[4\pi f_{i}(\hat{\bm\omega})\right]}\nonumber\\\nonumber\\&-&
\frac{1}{2}\sum_{i=1}^{N}
\int d\hat{\bm\omega}\int d\hat{\bm\omega}^{\prime}f_{i}(\hat{\bm\omega})
f_{i}(\hat{\bm\omega}^{\prime})
\sum_{j=1}^4\left\{P_2\left(\hat{\bm\omega}\cdot\hat{\bm\omega}'\right)+
\frac{\kappa}{\epsilon}\left[\left(\hat{\bm\omega}\times\hat{\bm\omega}'\right)\cdot\hat{\bm r}_j\right]
\left(\hat{\bm\omega}\cdot\hat{\bm\omega}'\right)
\right\}
\nonumber\\\nonumber\\&-&
\sum_{i=1}^{N-1}\int d\hat{\bm\omega}\int d\hat{\bm\omega}^{\prime}
f_{i}(\hat{\bm\omega})f_{i+1}(\hat{\bm\omega}^{\prime})
\left\{
P_2\left(\hat{\bm\omega}\cdot\hat{\bm\omega}'\right)+
\frac{\kappa}{\epsilon}\left[\left(\hat{\bm\omega}\times\hat{\bm\omega}'\right)\cdot\hat{\bm y}\right]
\left(\hat{\bm\omega}\cdot\hat{\bm\omega}'\right)
\right\}
\nonumber\\\nonumber\\&+&
\int d\hat{\bm\omega}f_{1}(\hat{\bm\omega})\Phi_1(\hat{\bm\omega})+
\int d\hat{\bm\omega}f_{N}(\hat{\bm\omega})\Phi_2(\hat{\bm\omega})
\label{free}
\end{eqnarray}
\end{widetext}
where $M$ is the number of particles in each plane, and $\{\hat{\bm r}_j\}$
are the four vectors $(\pm 1,0,0)$ and $(0,0,\pm 1)$ that connect a given particle to its four
neighbours in the $xz$ plane, and $\hat{\bm y}=(0,1,0)$ the unit vector along the $y$ axis (see Fig. \ref{fig1}). 

The orientational distribution functions $f_i(\hat{\bm\omega})$ were expanded in six basis functions $\Psi_{\alpha}(\hat{\bm\omega})$,
$\alpha=0,\cdots,5$:
\begin{eqnarray}
f_i(\hat{\bm\omega})=\sum_{\alpha=0}^5f_i^{(\alpha)}\Psi_{\alpha}(\hat{\bm\omega}).
\label{expan}
\end{eqnarray}
The functions $\Psi_{\alpha}(\hat{\bm\omega})$ correspond to real combinations of spherical harmonics with angular-momentum indices $l=0$ and $2$ and, therefore, are
orthonormal quadratic functions in $\hat{\bm\omega}$ ($\alpha=0$ corresponds to $l=0$ while $\alpha=1,\cdots,5$ correspond to combinations of the 
five spherical harmonics in the subspace $l=2$):
\begin{eqnarray}
\Psi_{0}(\hat{\bm\omega})&=&\frac{1}{\sqrt{4\pi}},\\\nonumber
\Psi_{1}(\hat{\bm\omega})&=&\sqrt{\frac{5}{4\pi}}P_2(\cos{\theta}),\\\nonumber
\Psi_{2}(\hat{\bm\omega})&=&\sqrt{\frac{15}{16\pi}}\sin{2\theta}\cos{\varphi},\\\nonumber
\Psi_{3}(\hat{\bm\omega})&=&\sqrt{\frac{15}{16\pi}}\sin{2\theta}\sin{\varphi},\\\nonumber
\Psi_{4}(\hat{\bm\omega})&=&\sqrt{\frac{15}{16\pi}}\sin^2{\theta}\cos{2\varphi},\\\nonumber
\Psi_{5}(\hat{\bm\omega})&=&\sqrt{\frac{15}{16\pi}}\sin^2{\theta}\sin{2\varphi}.
\label{expan1}
\end{eqnarray}
The fact that both particle-particle (\ref{interac}) and particle-plate interactions are quadratic in $\hat{\bm\omega}$ limits 
the (in principle) infinite series expansion of $f_i(\hat{\bm\omega})$ to six terms, cf. Eqn. (\ref{expan}) 
(the $l=1$ subspace is not included due to the head-tail symmetry of the particles --no terms linear in $\hat{\bm\omega}$ in the
potential). This means that the expansion (\ref{expan}) is complete and therefore it represents a full, parameterisation-free description of orientational order. 
Since $f_i(\hat{\bm\omega})$ is normalised to unity, $f_i^{(0)}=(4\pi)^{-1}$. The five remaining coefficients,
\begin{eqnarray}
f_i^{(\alpha)}=\int d\hat{\bm\omega}f_i(\hat{\bm\omega})\Psi_{\alpha}(\hat{\bm\omega}),\hspace{0.4cm}l=1,\cdots,5,
\label{OPs}
\end{eqnarray}
are a measure of orientational order in each plane $i$ in the lab frame. In fact, 
due to the symmetries of the confined system (the lowest-symmetry phase of which is the biaxial phase, with a distribution function
$f_i(\hat{\bm\omega})$ which is an even function of the angle $\varphi$), the coefficients $f_i^{(3)}$ and $f_i^{(5)}$ are zero, and consequently
only three order parameters, $f_i^{(1)}$, $f_i^{(2)}$ and $f_i^{(4)}$, are relevant. Using Eqn. (\ref{expan}) the free-energy functional 
(\ref{free}) can be expressed solely as a function of the set of $3N$ order parameters for the $N$ planes, 
$\{f_i^{(1)}, f_i^{(2)}, f_i^{(4)}\}$ (in order for this to possible in the case of the first, entropic term, the functions $f_i(\hat{\bm\omega})$ 
should be replaced by the expressions following from the equilibrium conditions $\delta F/\delta f_i=0$). The necessary angular functions were
computed using Gauss-Legendre quadratures. Minimisation of the free energy was performing using the conjugate-gradient method. 
The simple iterative (Picard) method on the equilibrium conditions $\delta F/\delta f_i=0$ was also used in some cases to check the results. 

The order parameters 
can most easily be interpreted in the proper reference frame $x'y'z'$, the one attached to the director. Because of symmetry, the director $z'$
lies always on the $xz$ plane, at an angle $\psi_i$ with respect to the $z$ axis. The rotation from the lab to the proper frame simply 
involves a rotation about the $y$ axis of angle $\psi_i$. From the rotational properties of spherical harmonics one obtains the relations
between the lab, $\{f_i^{(\alpha)}\}$, and proper, $\{\tilde{f}_i^{(\alpha)}\}$, order parameters. In the proper frame 
only the $\tilde{f}_i^{(1)}$ and $\tilde{f}_i^{(4)}$ order parameters are nonzero:
\begin{eqnarray}
\tilde{f}_i^{(1)}&=&P_2(\cos{\psi_i})f_i^{(1)} -\frac{\sqrt{3}}{2}\sin{2\psi_i} f_i^{(2)} + \frac{\sqrt{3}}{2}\sin^2{\varphi_i} f_i^{(4)},\nonumber\\
\tilde{f}_i^{(4)}&=&\frac{\sqrt{3}}{2}\sin^2{\psi_i}f_i^{(1)}+\frac{1}{2}\sin{2\psi_i} f_i^{(2)} + P_2(\cos{\psi_i})f_i^{(4)}.
\end{eqnarray}
$\tilde{f}_i^{(1)}$ is related with the standard uniaxial order parameter, $\eta_i=2\tilde{f}_i^{(1)}\sqrt{\pi/5}$,
which measures the uniaxial order of particles about the local director. The other parameter is related with the usual biaxial order parameter, 
$\sigma_i=4\tilde{f}_i^{(4)}\sqrt{\pi/15}$, which
accounts for possible departures from uniaxial symmetry of particles about the local director. The third and last relevant
parameter is the tilt angle $\psi_i$ (which also depends on the plane index, since in the cholesteric phase 
the director rotates from plane to plane). 

\begin{figure}
\includegraphics[width=0.9\linewidth,angle=0]{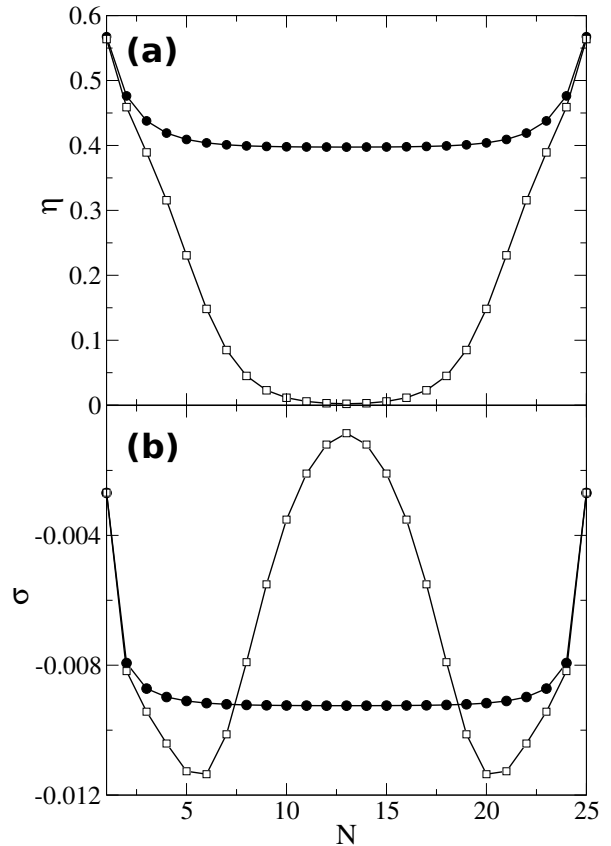}
\caption{Uniaxial (a) and biaxial order parameter (b) profiles of an isotropic (open squares) and 
a cholesteric (filled circles) phase coexisting in a symmetric slit pore with $N=25$. The tilt angle of the cholesteric
phase (not shown) rotates by approximately $2\pi$ across the cell. }
\label{fig3}
\end{figure}

\section{Results}

\subsection{Bulk system}

We first describe the bulk phase behaviour of the model. In the achiral system ($\kappa=0$) there is an isotropic-nematic phase transition
\cite{Fabbri,Zhang}. 
For any nonvanishing chirality ($\kappa\neq0$) the nematic phase is replaced by a cholesteric phase, and the system exhibits a first order isotropic-cholesteric (Iso-Cho) 
phase transition by decreasing the temperature. The uniaxial and biaxial order parameters as a function of the temperature for $\kappa/\epsilon=0.5$ are shown in Fig.~\ref{fig2}(a). 
The cholesteric pitch (not shown) is independent of the temperature, as can be demonstrated \cite{vanderMeer,Luckhurst}. The behaviour of the order parameters is qualitatively 
the same for any value of $\kappa$. 

In Fig.~\ref{fig2}(b) we represent the bulk phase diagram in the plane of chirality and temperature. The stronger the chirality is, the higher the 
phase transition temperature becomes. In the limit of vanishing chirality ($\kappa\rightarrow 0$), the transition temperature tends asymptotically to the isotropic-nematic transition 
temperature of the achiral system. 

The uniaxial and biaxial order parameters at the phase transition are depicted in Fig.~\ref{fig2}(c). As expected the biaxiality is very small. 
Finally, 
in Fig.~\ref{fig2}(d) 
we show the cholesteric pitch at the Iso-Cho phase transition as a function of $\kappa$. The cholesteric pitch increases rapidly for decreasing 
chirality, diverging ($p\rightarrow\infty$) in the limit $\kappa\rightarrow 0$. This limit corresponds to a system of achiral particles in which a nematic or, equivalently, 
a cholesteric with infinite pitch, is stable. The data closely follow the curve 
$p=4\pi(\arctan{\frac{2\kappa}{3\epsilon}})^{-1}$, an approximation that results by neglecting the entropy and minimising the internal 
energy with respect to the pitch.

\subsection{Confined system}

Next we confine the system in a symmetric pore made of two identical parallel plates. As mentioned before, we set the surface strength as
$\epsilon_s=\epsilon$. This choice corresponds to moderate or strong anchoring conditions on the two plates since, for relatively thin pores, the particle
orientation next to the plates never exceeds $2^{\circ}$ from the easy axis for stable configurations.  
Although we have studied different values of $\kappa$, here we only show results for $\kappa/\epsilon=1$, as no qualitative differences were found
for different values of the chirality strength. For this choice of $\kappa$ the cholesteric pitch is $p\simeq 21$ (in units of the 
lattice parameter).

\begin{figure}
\includegraphics[width=1.00\linewidth,angle=0]{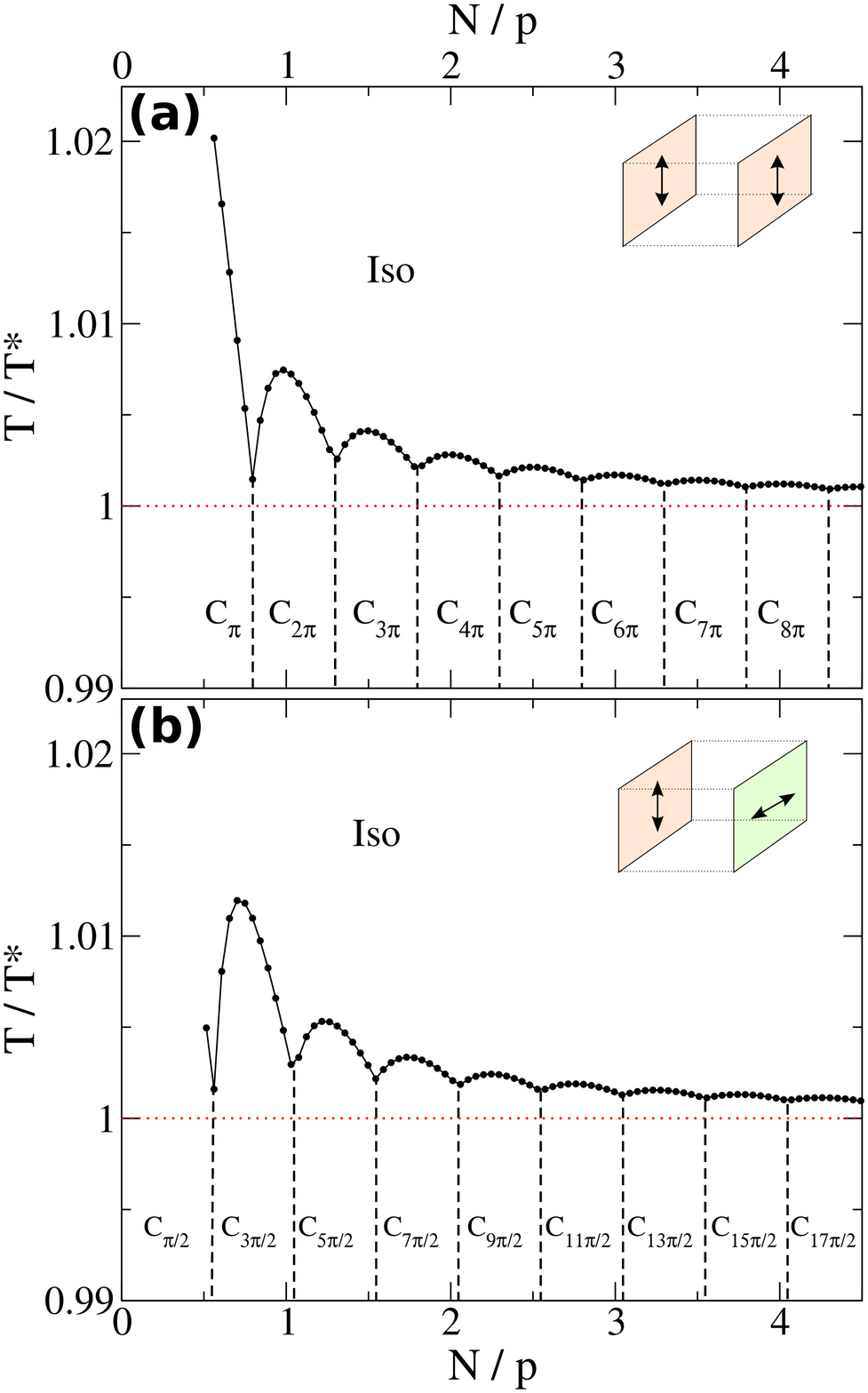}
\caption{Phase diagram of a confined cholesteric liquid crystal in (a) symmetric, and (b) asymmetric
pores in the scaled-temperature $T/T^*$ versus scaled pore-width $N/p$ plane. $T^*$ is the temperature for the bulk 
isotropic-cholesteric phase transition.
Filled circles indicate the isotropic-cholesteric binodal. 
Dashed lines represent the binodals along which two cholesteric phases with different number of helix indices coexist. The dotted red line indicates 
the temperature of the bulk transition.
In each case the geometry of the easy axes is depicted in the upper-right corner of the corresponding panel.}
\label{fig4}
\end{figure}

\subsubsection{Symmetric cell}

In the symmetric cell the anchoring directions are the same, i.e. $\hat{\bm n}=\hat{\bm m}$ (both vectors lying on the plane of the plates). 
In pores narrower than approximately half a cholesteric period ($N\lesssim p/2$), the capillary isotropic-cholesteric transition  
disappears, and the cholesteric grows from the isotropic state in a continuous fashion as temperature is decreased. 
The suppression of the capillary transition for narrow pores is typical of confined fluids. In a narrow pore 
surface interactions strongly affect the particle orientations in all layers, not just in those adjacent to the plates.  
Bulk orientational properties cannot compete with the strong anchoring imposed by the plates, and the absence of the bulk-surface
competition results in a perfectly continuous evolution of orientational order parameters, as dictated by surface potentials and temperature.
Despite the fact that there is no Iso-Cho phase transition, the system director, whose (symmetric) profile is close to linear, does show a total
twist mainly dictated by pore width and nearly independent of temperature. The total twist decreases as the pore becomes narrower.

In wider pores ($N\gtrsim p/2$) we found a first-order Iso-Cho phase transition. 
Order parameter profiles of the confined isotropic and cholesteric phases coexisting in a pore with $N=25$
are plotted in Fig.~\ref{fig3}. In the isotropic state the uniaxial order parameter, panel (a), vanishes in the middle of the pore, 
and it is relatively high close to the plates. In the cholesteric phase, the uniaxial order parameter is always higher than zero, 
and particles are also more ordered close to the plates. The biaxial order parameter, panel (b), is very small in the whole cavity. As can be seen,
it is always negative, meaning that particles fluctuate about the director with a larger amplitude in the direction of the helix (plate normal)
than in the direction of the plates.  

\begin{figure}
\includegraphics[width=0.85\linewidth,angle=0]{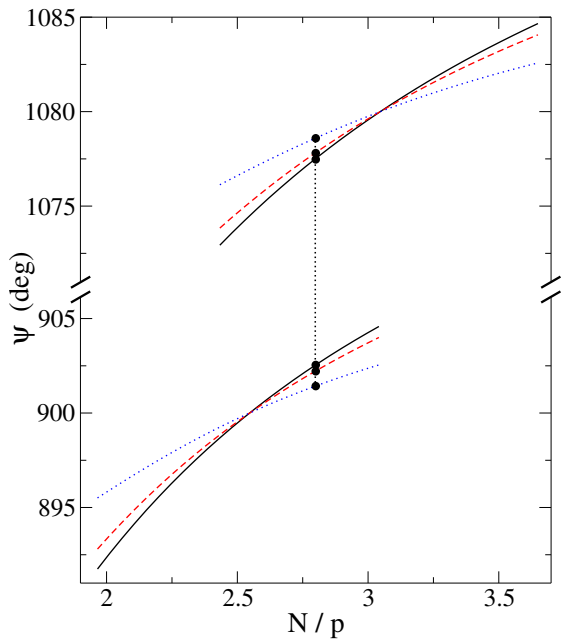}
\caption{Total twist $\psi$ (in degrees) of three cholesteric phases in a symmetric pore as a function of pore width scaled 
with cholesteric 
pitch, $N/p$. From bottom to top: $C_{5\pi}$ and $C_{6\pi}$ phases. In each case results are given for three different 
temperatures: continuous lines $T/T^*=0.800$ (where $T^*$ is the bulk transition temperature), dashed lines $T/T^*=0.880$, and
dotted lines $T/T^*=0.987$. Vertical dotted lines
correspond to the $C_{5\pi}-C_{6\pi}$ phase transition, whereas dots indicate the corresponding coexisting phases.}
\label{eltwist}
\end{figure}

The temperature at which the Iso-Cho transition occurs varies with pore size $N$. This is reflected in the phase diagram, depicted in 
Fig.~\ref{fig4}(a) in the pore size (scaled with the cholesteric pitch) versus temperature 
(scaled with the bulk Iso-Cho transition temperature) plane.
We first focus on the capillary Iso-Cho transition line. The transition is shifted to higher
temperatures with respect to the bulk; therefore, confinement promotes cholesteric order. 
The capillary line oscillates and tends asymptotically to the bulk transition temperature. The oscillatory behaviour is a direct 
consequence of a commensuration effect between the pore size and the cholesteric period. 
Local maxima of the capillary line occur when the elastic stress of the confined cholesteric is minimal, which happens 
if the pore size and the cholesteric period commensurate, i.e., $N/(p/2)\simeq k$, with $k$ a positive integer
(note that, due to the head-tail particle symmetry, the period of the cholesteric phase is in fact $p/2$, corresponding to half a turn of the 
director about the helix axis). Anyway the scaled-temperature range where this oscillatory structures takes place is rather small.

Besides the capillary Iso-Cho line, the system exhibits winding transitions in which two cholesterics with different 
periods coexist. The winding transitions occur always between cholesterics that differ in half a cholesteric period: 
$C_{\text{i}\pi}-C_{(\text{i+1})\pi}$, where the subscript $i\pi$, $i=1,2,3...$, denotes the total twist of the cholesteric 
helix from one plate to the other (save the small deviations of the director at the two plates arising from a large, but finite,
anchoring strength). At a winding transition, the cholesteric period of the coexisting $C_{\text{i}\pi}$ phase is expanded with respect 
to its equilibrium value, while the period of the coexisting $C_{(\text{i+1})\pi}$ phase is compressed. 
Hence, the size of the pore at which two cholesterics coexist is such that the elastic stress is maximum. Winding transitions merge 
with the Iso-Cho capillary line, giving rise to Iso$-C_{\text{i}\pi}-C_{(\text{i+1})\pi}$ triple points. 
The triple points are located on the local minima (low temperature) of the capillary Iso-Cho line. 

Note that winding transitions are almost 
vertical lines in the pore size-temperature phase diagram, implying that they are insensitive to temperature. This result can be traced 
back to the insensitivity of the cholesteric pitch with respect to temperature. 
To see this more clearly, Fig. \ref{eltwist} shows the total twist across the cell (from one plate to the other) 
of the director as a function of pore width $N/p$ for two phases, $C_{5\pi}$ and $C_{6\pi}$, in the region of the 
$C_{5\pi}-C_{6\pi}$ phase transition.
For each phase, three curves corresponding to three temperatures, $T/T^*=0.800$, $0.880$ and $0.987$ are shown. 
The three curves (and, in fact all curves corresponding to different temperatures) intersect at a point with a total twist 
equal to the number of half-periods, $5\pi=900^{\circ}$ for the $C_{5\pi}$ phase and $6\pi=1080^{\circ}$ for the 
$C_{6\pi}$ phase, and at a pore width slightly higher than the optimal value (which is given by $N=5(p/2)=2.5p$ and
$N=6(p/2)=3p$, respectively); this is due to the effect of the plates, which disappears as the pore width becomes larger. 
The $C_{5\pi}-C_{6\pi}$ transition is obtained when the free energies of the two structures become
equal. For all three temperatures, the transition is located at almost exactly the same pore width
(denoted by the dotted vertical line in the figure). In effect, due to the
temperature invariance of the pitch, the temperature plays a negligible role in driving winding transitions.
By contrast, a
model with temperature-dependent pitch would give rise to boundary lines presenting a
slope in the phase diagram.
Finally, as mentioned before, we note that, at coexistence, the phase with the smaller number of turns, $C_{5\pi}$, is expanded, whereas 
the one with the larger number of turns, $C_{6\pi}$, is compressed. This is because the transition point is located at wider (narrower) 
pores than optimal for the $C_{5\pi}$ ($C_{6\pi}$) phase.

\subsubsection{Asymmetric cell}

Next we confine the system in an asymmetric pore, in which plates favour perpendicular directions (easy axes satisfying 
$\hat{\bm n}\perp\hat{\bm m}$). All remaining parameters are set to the same values as in the previous symmetric case. 
Not surprisingly the phase diagram, shown in Fig.~\ref{fig4}(b), resembles the one of a symmetric pore. The only significant difference 
concerns the total twist of the confined cholesterics, which is always a multiple of $\pi/2$; therefore the cholesteric phases can be labelled as 
$C_{(2i+1)\pi/2}=C_{(i+\frac{1}{2})\pi}$, with $i=0,1,2...$. The subscript $(i+\frac{1}{2})\pi$ indicates the total twist inside of the pore. 
Again cholesteric phases differing in half a cholesteric period, $C_{(i+\frac{1}{2})\pi}-C_{(i+\frac{3}{2})\pi}$, coexist at the winding transitions.
The phase diagram is essentially shifted by half a period with respect to the symmetric case. However, 
the pore width below which the Iso-Cho capillary line disappears is very similar to that in the symmetric cell.

\section{Conclusions}

In this work we have calculated the phase diagram of a cholesteric liquid crystal confined in a slit pore between two parallel plates. 
To our knowledge this is
the first time the structure of winding transitions is explicitly obtained as a function of pore size and connected to the capillary
transition. We have investigated symmetric (identical plates) and asymmetric pores (easy axes of plates at an angle) pores.
As a result of confinement, the cholesteric phase is promoted with respect to the bulk, the binodal line exhibiting 
a oscillatory structure, and strong commensuration effects, giving rise to winding transitions, occur. The two types (capillary and
winding) of transitions are connected: the capillary line preempts the winding transition structure, and the oscillatory nature of the former
is due to the commensuration effects at work in the winding transitions. Our results pertain to the case where the pore size of the cell
is changed in the absence of an external field, a situation not common in practical applications; however, these effects have been
observed in Atomic-Force Microscopy experiments, where both pore size and temperature can be controlled. 

Finally, it is interesting to note that phase diagrams similar to the ones obtained here have been previously predicted in smectic 
\cite{PRL,PRE} or columnar \cite{Yuri} liquid crystals confined in slit pores using density functional theory. 
In these system layering transitions between states that differ in the number of smectic or columnar layers occur under confinement 
due to commensuration effects between the pore width and the layer spacing. Layering transitions play a role similar to 
the winding transitions analysed here; they are connected to the capillary isotropic-smectic or nematic-smectic lines,
a phenomenon which is the analogue of  
the connection between the winding transitions and the capillary isotropic-cholesteric line found for the cholesteric model. This analogy
results in a universal behaviour, stemming from the competition between an internal length and an imposed, external length.
However, some
differences exist. For example, the capillary transition lines at small pore widths is in some cases broken in the smectic case. Possible origins
of these differences (resulting from the theoretical approximations or from the interaction models) is currently under investigation.

\vspace{1cm}
\noindent{\bf Acknowledgements} 

This work was partially financed by the Portuguese Foundation for Science and Technology (FCT) through grant PTDC/FIS/119162/2010. 
E. V. also acknowledges support from MINECO (Spain) through grant FIS2013-47350-C5-1-R.

\end{document}